# Therapy Control and Patient Safety for Proton Therapy

*M. Grossmann*
Paul Scherrer Institut, Villigen, Switzerland

**Abstract**
This contribution describes general concepts for control and safety systems in proton therapy. These concepts are illustrated by concrete examples implemented in the Proscan facility at Paul Scherrer Institut (PSI).

## 1  Introduction

Protons have successfully been used for many years in the treatment of cancer. Their physical properties allow a better conformation of the dose to a target volume, which (depending on the tumour location) can give protons an advantage over conventional radiation therapy with photons.

When preparing a radiation therapy a therapy plan is created that designates the region in the patient to be treated and the dose to be applied. This plan is then translated into a steering file containing the machine information for the treatment facility. It is the task of the therapy control system to make sure that the treatment follows this prescription. The patient safety system supervises the treatment, detects dangerous situations and if necessary interrupts the treatment.

This article describes general concepts for proton therapy control and safety systems, illustrating them using examples from the proton therapy facility at Paul Scherrer Institute (PSI).

## 2  Proton therapy at PSI

PSI operates the Proscan facility for proton therapy [1]. The layout of the facility is shown in Fig. 1. The proton beam is produced by the superconducting 250 MeV cyclotron COMET. A degrader at the exit of the cyclotron reduces the beam energy according to the therapeutic requirements. The beam is then sent into four treatment rooms.

The OPTIS treatment room is dedicated to the treatment of ocular tumours (mostly melanoma). It uses a conventional scattering technique to apply the required dose. With this technique patients have been treated at PSI since 1984. The installation is an upgrade of the original facility and started operation in 2010.

Gantry 1 went into operation in 1996 as the world's first spot-scanning gantry that applies the dose by magnetic scanning of a focused pencil beam. It aims at the treatment of deep-seated tumours, many of them some kind of brain tumour.

With Gantry 2, PSI continued the technological development of magnetic pencil beam scanning [2]. It is based on the experience with Gantry 1 and is optimized for fast and advanced scanning modes, allowing the extension of treatment indications. Patient operations started in 2013.

The most recent treatment room Gantry 3 is currently under construction. Unlike the other rooms which were built by PSI, it is a commercial device based on Varian Medical System's ProBeam system. Treatment of patients is scheduled to start in 2017.

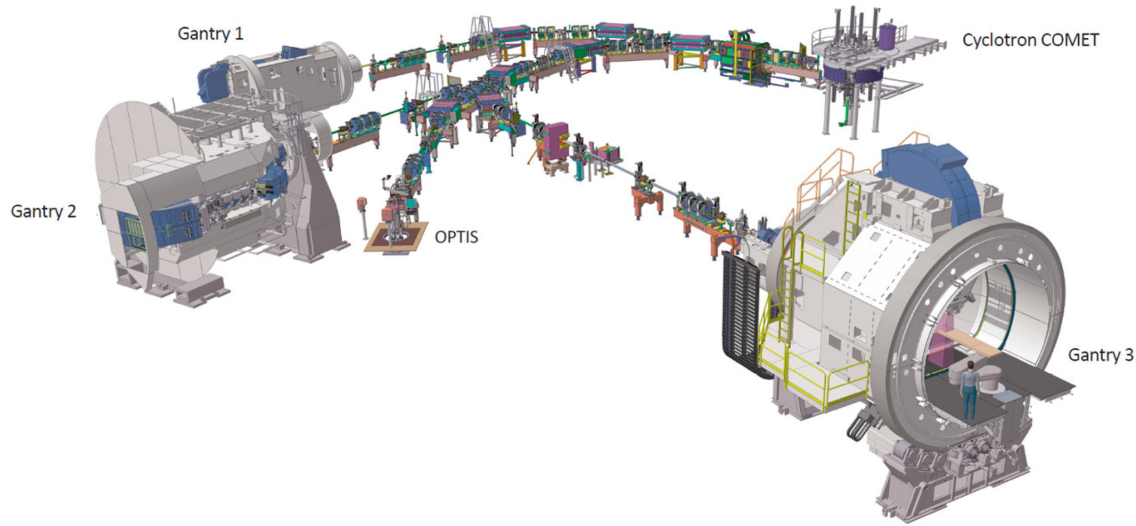

**Fig. 1:** Layout of the Proscan facility beamlines including the new Gantry 3 area. The existing treatment areas are Gantry 1, OPTIS 2 and Gantry 2.

## 3       Control systems

### 3.1     Machine control system

The *machine control system (MCS)* groups together all the subsystems necessary to control the accelerator and the beamline. Its purpose is to deliver a beam with modifiable beam energy, momentum spread and intensity to the treatment area.

The MCS can be based on a framework such as EPICS[1]. Access to the equipment under control is provided through *Versa Module Eurocards*, so-called *VME modules*, housed in several crates distributed over the Proscan facility. Each VME crate is equipped with a single-board computer called *input output controller (IOC)*. These IOCs can be accessed over the network using the EPICS concept of *channels* by other applications or user interfaces.

The MCS provides a *graphical user interface (GUI)* running on workstations located both in the central control room and at each treatment room. Write access from these workstations is restricted during therapy.

Communication between a treatment area and the MCS is organized through the *beam allocator gateway (BALL)*. Besides providing an interface to the MCS, BALL restricts write access to MCS-controlled devices in order to prevent interference from different areas during therapy.

### 3.2     Machine protection

The purpose of the machine protection system (at PSI called *run permit system (RPS)*) is to protect the beamlines and the accelerator from damage by the proton beam. It defines limits for the setting of devices in the beamlines, e.g. an upper and a lower limit for the current in magnets along the beam line. If the current setting is outside of these limits, the beam cannot be switched on. If the beam is on when a violation of the supervised limits occurs, it is switched off. The RPS has its own independent interface to a set of final elements for this purpose.

---

[1] EPICS is a software framework for experimental physics and industrial control systems.

The limits used by the RPS depend on the treatment area that requests the beam. The RPS is informed by the MCS which treatment area is requesting the beam and then sets the limits accordingly.

## 3.3 The concept of mastership

During treatment of a patient only one area can have full control over the facility, and no other area is able to interfere. Based on the planned therapy flow, one area requests exclusive access, called *mastership*, to the beamline from BALL. If available, BALL grants mastership, giving exclusive access to both the shared and local beamline sections all the way upstream to the cyclotron. This includes control of the degrader (energy selection) and the kicker magnet (beam on/off control). The requesting area becomes 'master' and the patient treatment scheduled in this area can be performed.

## 3.4 Therapy control system

The *therapy control system (TCS)* consists of two components, the *therapy delivery system (TDS)* and the *therapy verification system (TVS)*.

TDS and TVS run on IOCs in separate VME crates. These crates contain (besides the IOCs) the necessary modules to interface with the actuators and sensors used to perform the treatments, and with the other control system parts (see Fig. 2).

Proceeding spot by spot, the TDS initiates the setting of the actuators to control the beam position (the two sweeper magnets, the patient table, the gantry motors and the beamline) and sets the required dose for the spot to be applied. When all preparations are successfully completed, the command to switch on the beam is issued.

The TVS checks that the spot sequence proceeds as planned. It verifies the correct setting of the actuators by means of dedicated sensors (e.g. Hall probes to check the field of the sweeper magnets; position encoders to check the positions of the patient table and gantry body). In addition, it performs a number of checks before, during and after the spot application (e.g. verification of the dose applied and of the spot position).

In case of a failure, the TDS and TVS can interrupt the treatment by generating an interlock. Both TDS and TVS use individual steering files which contain definitions of the dose and of the nominal values for actuators and sensors for each spot.

The user interface to the TCS consists of two applications: the *GUI server*, communicating with TDS and TVS over the network, and the *GUI client*, communicating with the GUI server over the network. There may be more than one instance of the GUI client running at a time. The GUI server guarantees that only one GUI client has write access to the TCS.

The proton dose is monitored by ionization chambers. The primary (dose-defining) monitor is located just upstream of the patient. Additional monitors in the treatment room and along the beamline supervise the correct function of the primary monitor and serve for additional safety functions (e.g. interrupting the treatment when beam intensity becomes too high).

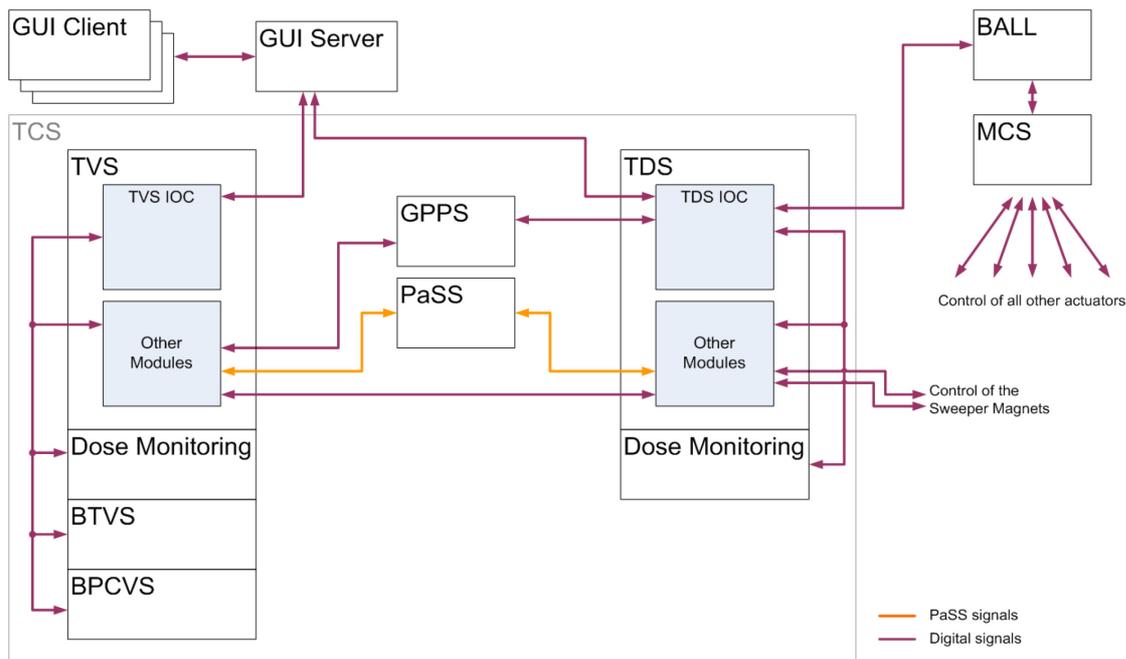

**Fig. 2:** Simplified scheme of the therapy control system (TCS) with its interface to associated control systems and the GUI server and clients. PaSS signals are failsafe connections (three-wire current loop), digital signals are I/O signals like TTL, or serial, parallel and network connections.

## 4 Patient safety system

### 4.1 Principle

The purpose of the *patient safety system (PaSS)* is to protect the patient from radiation hazards by minimizing the risk for an uncontrolled irradiation.

The PaSS follows the principles of any safety system:

- *sensors* measure critical parameters;
- a *logic* evaluates the data provided by the sensors and decides if the ongoing (therapy) process should be interrupted;
- *final elements* reach the safe state (no beam, stop of mechanical motion).

The PaSS is implemented with logic hard coded in electronic chips and using only point-to-point connections, avoiding the use of communication buses. Reflecting the architecture of the Proscan facility, the PaSS has local and central parts (schematically shown in Fig. 3):

- the local PaSS for a specific treatment area, with its own sensors, logic and final elements;
- the central PaSS for those parts of the facility that are used by all treatment areas, with sensors, logic and final elements that must be shared between treatment areas.

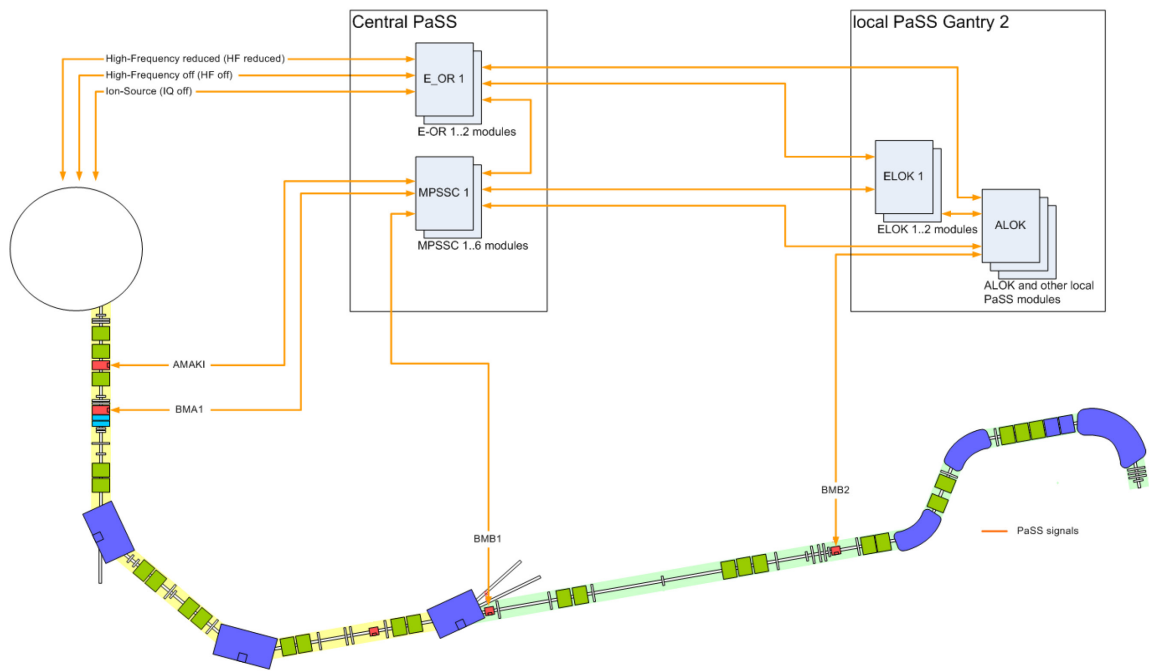

**Fig. 3:** General architecture of the patient safety system with local and central components, each containing several modules. This figure is largely simplified: the single wires shown in the figure are in reality multiple hardware, point-to-point connections. Inputs to the PaSS (from sensors directly and from control systems) are not shown.

### 4.2 Final elements

Several final elements allow us to switch off the beam to prevent uncontrolled irradiation. Both redundancy and diversity are required. At PSI the following elements are used (see also Fig. 3 for their location).

#### 4.2.1 Kicker magnet AMAKI

The kicker magnet AMAKI is the fastest device dedicated to turning the beam on and off. If the magnet is energized, it deflects the beam into a beam dump. AMAKI was primarily designed as a steering element for spot scanning. Due to its fast reaction time and its ability to handle many on/off cycles in an extremely reliable way, it also has the function of a safety element. A magnetic switch is used to supervise the magnetic field of the AMAKI. Reaction time for AMAKI is 300 µs.

#### 4.2.2 High-frequency generator

The high-frequency (HF) generator of the cyclotron can be used as final element in two stages: the power can be reduced to 80% (HF_RED)—this already stops acceleration of protons completely but prevents cooling down of the generator, which can lead to beam instabilities. In a second stage the generator can be completely switched off (HF_OFF). Reaction time for both stages is 400 µs.

#### 4.2.3 Ion source

Powering off the ion source at the centre of the cyclotron prevents protons entering the accelerator (IQ_OFF). Reaction time for the ion source is 20 ms.

### *4.2.4    Local fast mechanical blocker*

At the entrance point of the proton beam into the treatment room a fast mechanical beam blocker (BMB2) is installed. It is a block of 10 cm copper which can be moved in or out of the beamline by pressurized air and a mechanical spring. In case of failure of the pressurized air it will fall into the beam-blocking position by gravity. Reaction time for the local fast mechanical blocker is 60 ms.

### *4.2.5    Local slow mechanical blocker*

Located in the local part of the beam line, but still upstream of the treatment room, is a slow mechanical blocker (BMB1) made of graphite. It can be moved in or out of the beam line by pressurized air. In case of failure of the pressurized air it will fall into the beam-blocking position by gravity. Reaction time for the local slow mechanical blocker is 1 s.

### *4.2.6    Central slow mechanical blocker*

Located near the exit of the cyclotron is a slow mechanical blocker (BMA1) made of graphite. It can be moved in or out of the beam line by pressurized air. In case of failure of the pressurized air it will fall into the beam-blocking position by gravity. Reaction time for the central slow mechanical blocker is 1 s.

## 4.3    The beam switch-off functions

The beam *switch-off* can have two different functions.

- A *control function*: the TCS controlling the dose application switches the beam off when the specified dose has been applied.
- A *safety function*: a potential radiation hazard is detected and the PaSS switches the beam off. Depending on the risk associated with the radiation hazard and the source of the hazard, three different interlock levels have been defined. They are labelled ALOK, ATOT and ETOT (see Fig. 4)

### *4.3.1    Beam switch-off control function*

The beam *switch-off* control function is used during regular operation to terminate the spot once the nominal dose has been reached. At PSI this is realized by energizing the kicker magnet AMAKI, the *primary final element*.

Since in spot scanning the dead time between the single dose element applications needs to be kept below a few milliseconds, the regular beam *switch-off* control function does not trigger any mechanical beam blockers to be closed.

### *4.3.2    ALOK beam switch-off safety function*

If a problem with any equipment of the local beamline occurs which might cause a radiation hazard, an ALOK is raised. An example is an unexpected beam position.

In the presence of an ALOK, the safe state, i.e. beam off, is achieved by the activation of the following final elements:

- the fast local beam blocker is closed (BMB2);
- the kicker magnet is energized (AMAKI).

The correct reaction to an ALOK is supervised. If one or several monitoring functions do not respond within the defined timeout, the ALOK is escalated to the next interlock level ATOT. Reasons which can lead to an escalation are:

- AMAKI status not changed after timeout;

- BMB2 not closed after timeout;
- nozzle dose monitors measure beam after timeout.

### 4.3.3  *ATOT beam switch-off safety function*

If a problem originating from a device in the shared beamline occurs and could cause a radiation hazard or if an ALOK was escalated, an ATOT is raised. This happens for example when, after a beam *switch-off* command from the TDS, the beam current does not reach zero within the specified time limit.

In the presence of an ATOT, the following final elements are activated:

- the local slow beam blocker is closed (BMB1);
- the central slow beam blocker is closed (BMA1);
- the cyclotron HF generator power is reduced to 80% (HF_RED).

The correct reaction to an ATOT is supervised. If one or several monitoring functions do not respond within the defined timeout, the ATOT is escalated to the next interlock level ETOT. Reasons which can lead to an escalation are:

- BMA1 not closed after timeout;
- BMB1 not closed after timeout;
- HF power not reduced after timeout;
- beam line dose monitors measure beam after timeout.

### 4.3.4  *ETOT beam switch-off safety function*

In case of a high-risk radiation hazard, the ETOT (emergency interlock) signal is raised. This happens when:

- a person operates a mechanical emergency-stop button; or
- beam is detected in a treatment area which is not master; or
- an ATOT was escalated.

In the presence of an ETOT, the following final elements are activated:

- the cyclotron HF generator is powered off completely (HF_OFF);
- the cyclotron ion-source power supply is powered off (IQ_OFF).

| Final Element | Control Function | Safety Function | | | PaSS Components |
|---|---|---|---|---|---|
| BMB2 | | ALOK | | | local PaSS |
| AMAKI | beam off | | escalates | | |
| BMB1 | | | ATOT | | MPSSC |
| BMA1 | | | | | |
| HF reduced | | | | escalates | |
| HF off | | | | ETOT | E_OR |
| IQ off | | | | | |

**Fig. 4:** Summary of the final elements used by the regular beam-off command and by each of the safety functions, ALOK, ATOT and ETOT. The column to the right shows which of the PaSS components is responsible for controlling the respective final element.

### 4.4 Shared and redundant access to the common final element subsystems

With the exception of BMB2, access to the final elements is shared between all treatment areas. This access is implemented by the two components of the central PaSS:

- the *main patient safety switch and controller (MPSSC)*; and
- the *emergency-or module (E-OR)*.

    The MPSSC implements the following functions:

- area reservation and mastership validation:
    - only one single slow local beam blocker BMB1 can be open at a time and this only for the master area;
    - the mastership setting defined by BALL is cross-checked with direct hardware signals from the TCS of the treatment areas;
- ALOK safety function and AMAKI control:
    - the ALOK function is enabled for the master area exclusively;
    - access to the regular *beam on/off* command is restricted to the master area exclusively;
- ATOT safety function:
    - the ATOT safety function is restricted to the master area exclusively.

    Redundantly, the E-OR module implements the ETOT safety functions:

- HF_OFF;
- IQ_OFF.

    The E-OR allows all treatment areas to trigger the ETOT function at all times.

## 4.5 Implementation

The PaSS logic is implemented on several XILINX Spartan 2 FPGAs[2] (Fig. 5). These FPGAs are placed on industry-pack (IP) modules. Each module provides 10 channels (input and output). Hytec 8003 VME carrier boards can hold up to four IP modules. The carrier boards provide power to the IP modules and an interface to perform resets, set modes and read out the status of the IP modules. A transition board on the back side of the VME crate provides connections from the sensors and to the final elements.

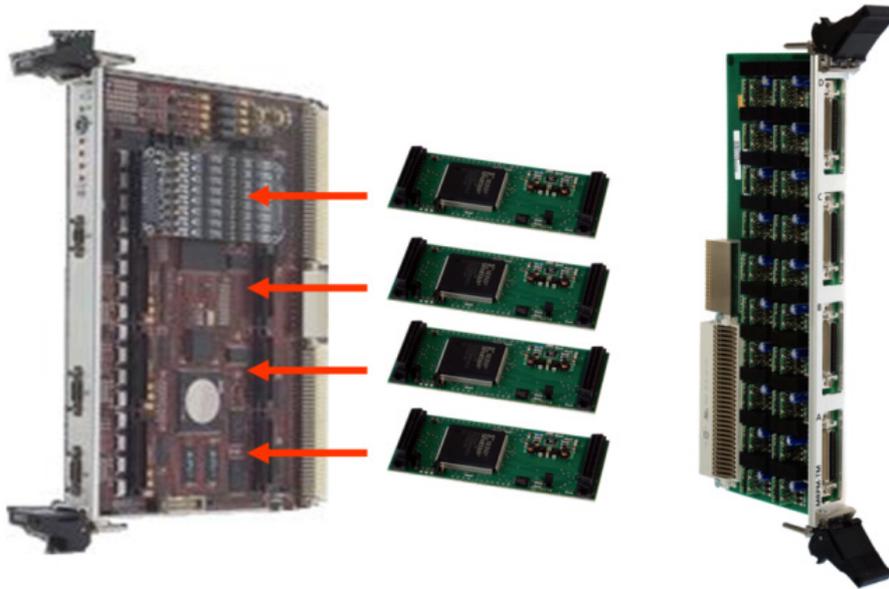

**Fig. 5**: Hardware implementation of the PSI patient safety system: four IP modules are mounted on a VME carrier board; the transition module on the right provides connections to sensors and final elements.

## 5 Integration of the commercial Gantry 3

Gantry 3 [3] is the latest extension of the PSI proton therapy facility. As already mentioned in the introduction the design is not based on PSI technology but the gantry is built by Varian Medical Systems.

Obviously Gantry 3 is delivered with its own control system. The challenge for the control system integration is to merge two quite different worlds: on one hand Varian's ProBeam scanning system, on the other hand PSI's existing systems, most of them built in-house. The Varian scanning system needs access to the cyclotron, the beamlines and the safety elements to control the beam and to switch it off in case of an interlock.

The chosen architecture leaves most of the existing systems untouched, with newly developed interfaces to connect them; see Fig. 6. The Varian system controls the scanning system (scanning magnets and dose monitors), the gantry and patient table positioning and the part of the beamline on the gantry itself. PSI systems control the upstream part of the beamline, the cyclotron and the central components of the PSI safety system.

The PSI part of the interface consists of two components, the *TCS* and the *PaS*S adapters.

---

[2] A field-programmable gate array (FPGA) is an integrated circuit designed to be configured by a customer or a designer

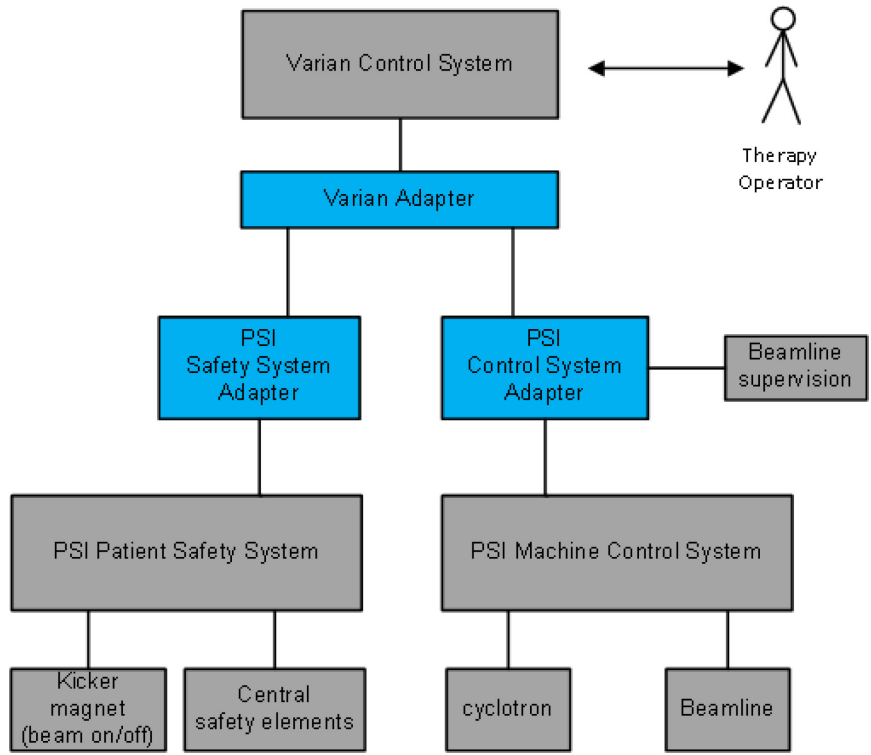

**Fig. 6**: Integration of Gantry 3 into PSI's control and safety systems is provided by the 'adapter' interfaces

## 5.1 TCS adapter

The TCS adapter communicates with the Varian system over a network connection. It serves as a gateway to PSI's machine control system. Typical commands are requests for treatment room reservation, beam energy and intensity. It also supervises the correct setting of the energy-selection system. It is implemented as a VME system with a Motorola single-board computer running VxWorks.

## 5.2 PaSS adapter

The PaSS adapter constitutes the interface between the Varian (local) and PSI (central) systems. While re-using most of the technology from the existing treatment rooms it was decided to program the logic on a state-of-the-art platform [4]. The choice fell for the IFC1210 controller, developed jointly by PSI and IOxOS [5]. It features a user-programmable Virtex 6 FPGA chip. It contains the safety system logic which after system startup is totally autonomous. On board are a further two Power PC CPUs running SMP Linux. They provide a standardized EPICS communication interface. This is used by the GUI, which provides access to the safety logic and automated actions like logging and statistics.

A generic platform, the so-called *signal converter box (SCB)*, supports the interconnection of IO signals from and to all subsystems. Data exchange between the SCB and the logic controller is handled over a high-speed communication link. It was developed jointly by PSI and Supercomputing Systems (SCS) [6].

For the development of the safety logic the same process as for the existing areas is followed. It comprises extensive verification and validation steps to ensure the correctness and integrity of the logic.

**Acknowledgements**

The development of PSI's control and safety systems and equally the whole proton therapy facility was great team work. The author would like to thank the members of the Center for Proton Therapy and the many others at PSI who contributed to this effort.

**References**

[1] J.M. Schippers *et al.*, *Nucl. Instrum. Methods* B **261** (2007) 773. http://dx.doi.org/10.1016/j.nimb.2007.04.052

[2] E. Pedroni *et al.*, *Z. Med. Phys.* **14** (2004) 25, http://dx.doi.org/10.1078/0939-3889-00194

[3] A. Koschik *et al.*, PSI Gantry 3: Integration of a new gantry into an existing roton therapy facility, Proc. IPAC16, 2016. http://jacow.org/ipac2016/papers/tupoy014.pdf

[4] P. Fernandez *et al.*, Reusable patient safety system framework for the proton therapy center at PSI, Proc. ICALEPCS2015, Pre-Press Release, 2015.

[5] IOxOS Technologies, IFC_1210 – Intelligent FPGA controller P2020 VME64x single board computer, rev. A0, data sheet, 2011, http://www.ioxos.ch

[6] Supercomputing Systems, http://www.scs.ch